\documentclass[aps,pre,print,twocolumn, footinbib,superscriptaddress]{revtex4-2}

\usepackage{microtype}
\usepackage{amssymb,amsmath}
\usepackage{graphicx}
\usepackage{color}
\usepackage{listings}
\usepackage{float}
\usepackage{wrapfig}
\usepackage{amsmath}
\usepackage{gensymb}
\usepackage{soul}
\usepackage{chemmacros}

\usepackage{upgreek}
\usepackage{enumerate} 
\usepackage[linkcolor = blue, citecolor = blue, urlcolor = blue, colorlinks = true]{hyperref}

\usepackage{commath,amssymb}
\usepackage{ushort}
\usepackage{multirow}
\usepackage{cleveref}
\usepackage{epstopdf}

\crefname{equation}{Eq.}{Eqs.}
\crefname{figure}{Fig.}{Figs.}

\usepackage[separate-uncertainty]{siunitx}

\usepackage{hyphenat}
\usepackage{booktabs}

\usepackage{tabularx}
\usepackage{array}
\usepackage{makecell}
\setcitestyle{super,open={},close={}}

\newcommand{\affleiden}{\affiliation{\small Leiden Institute Of Physics, Leiden University, P.O. Box 9504, 2300 RA Leiden, The Netherlands}}

\mathcode`\,="213B 

\begin{document}

\title{{\Large\bf Soft and stiff normal modes in floppy colloidal square lattices}}

\author{Julio Melio}
\affleiden
\author{Silke E. Henkes}
\affleiden
\author{Daniela J. Kraft}
\email[Corresponding email: ]{kraft@physics.leidenuniv.nl}
\affleiden

\date{\today}

\begin{abstract}
Floppy microscale spring networks are widely studied in theory and simulations, but no well-controlled experimental system currently exists. 
Here, we show that square lattices consisting of colloid-supported lipid bilayers functionalized with DNA linkers act as microscale floppy spring networks. We extract their normal modes by inverting the particle displacement correlation matrix, showing the emergence of a spectrum of soft modes with low effective stiffness in addition to stiff modes that derive from linker interactions. Evaluation of the softest mode, a uniform shear mode, reveals that shear stiffness decreases with lattice size. Experiments match well with Brownian particle simulations and we develop a theoretical description based on mapping interactions onto linear response to describe the modes. Our results reveal the importance of entropic steric effects, and can be used for developing reconfigurable materials at the colloidal length scale. 

\end{abstract}

\maketitle

Networks below the mechanical stability threshold prescribed by constraint counting exhibit large-scale, low energy structural deformation modes \cite{Lubensky2015}. These floppy modes play an important role in a wide range of systems ranging from elastic networks \cite{Ellenbroek2015, Chen2022} to particle packings \cite{Liu2010}. They are also important to understanding the rheological behavior of colloidal gels \cite{Rocklin2021}, the glass transition in disordered solids \cite{Widmer-Cooper2008}, and protein flexibility \cite{Meireles2011}. 
Both Maxwell constraint counting as used for frictionless ordered \cite{Lubensky2015} and disordered \cite{van2009jamming} packings and networks, and more sophisticated approaches like rigidity percolation \cite{Ellenbroek2015} predict that floppy modes are strictly zero energy deformations in linear response. However, floppy modes in microscale networks have been predicted by simulations and theoretical arguments to be stabilized by thermal fluctuations, resulting in nonzero shear moduli\cite{Mao2015,Dennison2013,Zhang2016, hu_2018}. 

Using microscale building blocks such as colloidal particles to make floppy structures would allow testing of these predictions and lead to a better understanding of thermally excited floppy modes, and in turn, exploiting floppy modes in mechanical metamaterials to obtain specifically engineered capabilities such as shape morphing \cite{Coulais2016} or topology dependent mechanical properties \cite{Xiu2022}. While colloidal structures with rearrangements have been experimentally realized using colloids coated with surface-bound DNA linkers which can roll over each others surface close to the melting temperature\cite{Wang2015, fang2020} and patchy particles\cite{mao2013}, neither allows for fixed bond network topology and are thus limited to structures accessible by free energy minimization. An experimental model system that combines full flexibility with a fixed bond network topology is currently lacking. 
 
Here we present an experimental system that behaves as a floppy microscale spring network and allows \textit{in situ} observation of the dynamics and thermal excitations by optical microscopy. We build two-dimensional square lattices from colloid-supported lipid bilayers (CSLBs) and find that they display soft shear modes with a small but finite stiffness 2-3 orders of magnitude softer than the stiffer compression modes, which is different from disordered \cite{Rocklin2021} or large crystalline \cite{Keim2004,Chen2013} systems. All vibrational modes could be extracted from the positional data and showed good agreement to Brownian particle simulations. Furthermore, a theoretical description based on mapping interactions onto linear response matches well with experiments and simulations. Our results demonstrate that floppy colloidal structures are a powerful model system to test predictions on thermal floppy networks by direct observation of the particle positions.

\begin{figure*} 
\includegraphics[width=\textwidth]{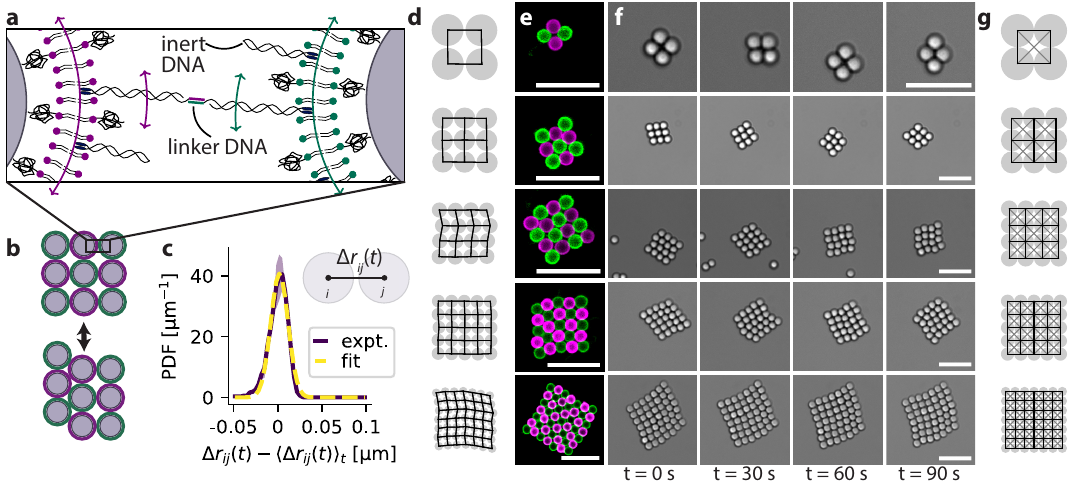}
\caption{\textbf{Flexible colloidal square lattices.} \textbf{a)} A schematic overview of the flexible binding mechanism, in which silica particles are coated by a fluid lipid bilayer that is functionalized with laterally mobile PEGylated lipids, inert DNA, and linker DNA. \textbf{b)} Schematic example of a floppy mode deformation in a 3x3 square lattice. \textbf{c)} The distribution of the distance between two particles $i$ and $j$, $\Delta \vec{r}_{ij} - \langle\Delta \vec{r}_{ij}\rangle_\mathrm{t}$, measured on particle pairs can be well-approximated with a normal distribution yielding a bond stiffness of $k_{\mathrm{bond}} =$ 43 $\pm$ $1$ $\mu \mathrm{N m^{-1}}$. \textbf{d)} Time average particle positions as measured in experiments, where the bond network is indicated in black. \textbf{e)} Confocal images of the $\mathrm{n\times n}$ square lattices with $\mathrm{n = 2,3,4,5,7}$ where the color indicates functionalization with different, complementary DNA linkers. \textbf{f)} Brightfield snapshots of the same lattices taken 30 s apart. \textbf{g)} Spring networks used in the theoretical description with the particle bonds indicated in black and the effective diagonal interactions indicated in grey. Scale bars are 10 $\mathrm{\mu m}$.
\label{LOfig:fig1}} \end{figure*}

\textbf{Creating experimental floppy colloidal square lattices}. To create floppy square lattices, we employ colloid supported lipid bilayers (CSLBs) functionalized with two complementary strands of DNA linkers as described in Verweij \textit{et al.} \cite{Verweij2023}. Briefly, silica particles with diameter $d$ = (2.12 $\pm$ 0.06) $\mathrm{\mu m}$ are coated with a fluorescently labeled lipid bilayer by the addition of small unilamellar vesicles (SUVs). The thus obtained CSLBs are functionalized with complementary DNA strands with 11 base pair long sticky ends, resulting in the ability to selectively form bonds several $k_\text{B}T$ strong, as shown in Fig. \ref{LOfig:fig1}a. The DNA can freely move within the bilayer, which allows the particles to fully reconfigure with respect to each other while remaining bound and thereby imparting flexibility onto the structures. Fluorescent lipids, visible in the confocal images in Fig. \ref{LOfig:fig1}e, are used to distinguish the functionalization of the particles. 

We assemble the particles into square lattices of size $n \times n$ with $n = 2,3,4,5,7$, using optical tweezers. We chose square lattices as an easily scalable floppy system: Square lattices, which are isostatic in the infinite limit using simple constraint counting \cite{Maxwell1864}, in fact admit $2n$ floppy modes that correspond to shear deformations between layers when cut to a finite size \cite{Lubensky2015}. An example of such a floppy mode displacement for a 3x3 lattice is schematically shown in Fig. \ref{LOfig:fig1}b. In addition, once assembled from complementary CSLBs, the square lattice structure cannot change its network topology because bonds between particles diagonally opposite of each other are not possible by design.  

We observe the dynamics of the assembled lattices with a brightfield microscope at a frame rate of 20 fps. The particles move predominantly in 2D (with a gravitational height of 53 $\mathrm{nm}$) and have previously been found to be freely jointed to very good approximation.\cite{Verweij2021, Verweij2023} Because of their size, the thus assembled lattices undergo thermal motion in the form of translation and rotation but also by changing their conformations. The large flexibility of our square lattices is evident from the snapshots taken 30 s apart shown in Fig. \ref{LOfig:fig1}f and Supplementary Movies 1-5, where the thermally excited shear modes are clearly visible. The bonds between the particles are very strong as they consist of many DNA linkers and thus - while allowing rearrangements, even at room temperature - do not break unless the temperature is elevated close to the melting temperature. This fixed topology of the bond network in combination with full flexibility makes this system an ideal choice for studying microscopic spring networks. 

We start by showing that the multivalent DNA bonds between the CSLBs can be approximated by harmonic springs. To do so, we quantify the particle positions of a pair of particles using Holopy.\cite{Barkley2020, Verweij2023} The sub-pixel accuracy of the particle tracking detects variations on the order of nanometers, and thus allows us to track not only the floppy modes with high accuracy but also elastic normal modes which correspond to smaller displacements involving a change in distance between particles, on the scale of several tens of nanometers. We plot the bond length variation $\Delta r_{ij} - \langle\Delta r_{ij}\rangle_\mathrm{t}$ for two bound particles in Fig. \ref{LOfig:fig1}c, where $\Delta r_{ij}$ is the distance between the centers of particles $i$ and $j$, and $\langle\dots\rangle_t$ denotes an average over the measurement time. A Gaussian fit with spring constant $k_{\mathrm{bond}} =$ $43\pm 1$ $\mathrm{\mu N m^{-1}}$ agrees well with the experimental data. The multivalent patch of DNA linkers therefore acts as an effective spring implying that CSLB based colloidal structures are effectively microscale spring networks, at least to lowest order. The network structures for the different lattice sizes are schematically drawn in Fig. \ref{LOfig:fig1}d, where the springs are indicated in black.

\begin{figure*} 
\centering     
\includegraphics[width=\textwidth]{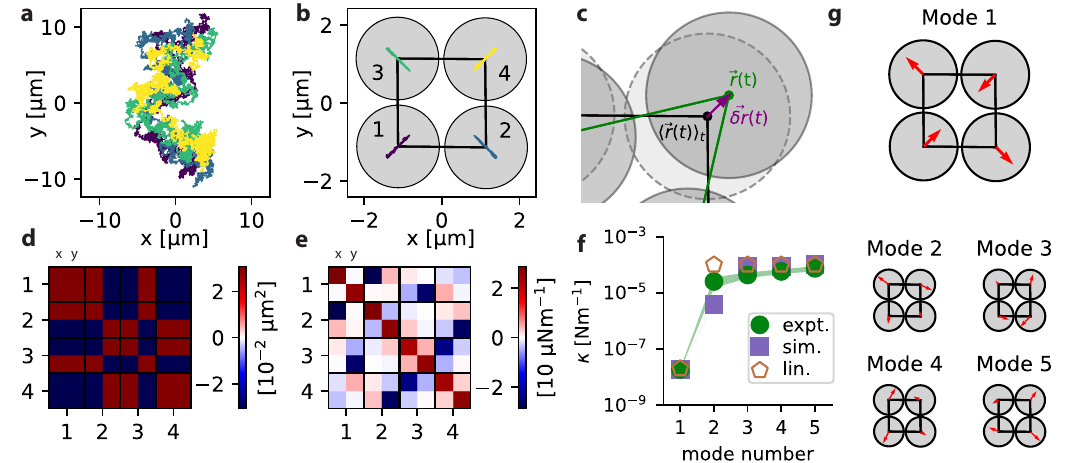}
    \caption{\textbf{Mode calculation for a 2x2 structure.} \textbf{a)} A typical trajectory for a 2x2 lattice, where the different colors represent the four different particles. \textbf{b)} The same trajectory after subtracting translation and rotation, where the average position $\langle\vec{r}(t)\rangle_t$ is shown and the particles are labeled. \textbf{c)} A schematic showing the equilibrium position $\langle\vec{r}(t)\rangle_t$, the position at time $t$, $\vec{r}(t)$, and the displacement vector $\vec{\delta r}(t)$. \textbf{d)} The covariance matrix of the displacement vectors $\mathbf{C_p}$, which is pseudo-inverted to obtain \textbf{e)} the stiffness matrix $\mathbf{K}$. \textbf{f)} The eigenvalues of $\mathbf{K}$ representing the mode stiffnesses \(\kappa\) for experiments (expt.), Brownian particle simulations (sim.) with $k_\mathrm{bond}=$ 50 $\mathrm{\mu Nm^{-1}}$, and linear response (lin.) with $k_\mathrm{bond}=$ 50 $\mathrm{\mu Nm^{-1}}$ and $k_\mathrm{diag.}=$ 10 $\mathrm{n Nm^{-1}}$. \textbf{g)} The eigenvectors of $\mathbf{K}$ respresenting the vibrational modes. \label{LOfig:fig2}} \end{figure*}

\textbf{Results and interpretation.} We start by analyzing the particle trajectories to obtain the vibrational modes for a 2x2 lattice with $N=n^2=4$ particles and $2N=8$ degrees of freedom. A typical trajectory of particles is shown in Fig. \ref{LOfig:fig2}a. As the system freely translates and rotates in $2D$, we first subtract the motion due to these $3$ trivial global degrees of freedom to obtain the trajectory in Fig. \ref{LOfig:fig2}b (see SI section I\cite{supplementary} for details). As expected, particles move predominantly on the diagonals, corresponding to the shear-like motion observed in experiments. 

For colloidal glasses\cite{chen2010low,kaya2010normal,ghosh2010density} and crystals\cite{Keim2004,Chen2013}, which do not have floppy modes, equilibrium fluctuation-dissipation relations have been exploited to link particle fluctuations to the dynamical matrix, in a spirit similar to passive microrheology \cite{mizuno2008active}. We use this fluctuation-inversion method based on equipartition in the form developed by one of us \cite{Henkes12}.
Briefly, we compute the particle displacement vector $\delta \vec{r}(t) = \vec{r}(t) - \langle\vec{r}(t)\rangle_t$, where $\langle\dots\rangle_t$ denotes a time average over the experiment. From this, the $2N\times 2N$ covariance matrix of the displacements, $\mathbf{C_p}$, is calculated. In the long-time limit and in linear response, the stiffness matrix \(\mathbf{K}\) is related to $\mathbf{C_p}$ by $\mathbf{K} = k_\mathrm{B} T \mathbf{C_p^{-1}}$ \cite{Henkes12}, where we now need to use the pseudo-inverse because $\mathbf{C_p}$ is singular \cite{Penrose1956} as a result of subtracting the trivial modes. The eigenvectors of $\mathbf{C_p}$ and $\mathbf{K}$ represent the normal modes and the eigenvalues of $\mathbf{K}$, $\kappa$, represent the mode stiffnesses. We show the covariance $\mathbf{C_p}$ and stiffness $\mathbf{K}$ matrices in Figs. \ref{LOfig:fig2}d and \ref{LOfig:fig2}e, respectively. $\mathbf{C_p}$ only contains two distinctive values
matching the earlier observation that particles move on diagonals.
In $\mathbf{K}$, in addition to the diagonal terms that indicate confinement, we observe large stiffness values along the pair bond directions, as well as smaller values that indicate next-nearest-neighbour (NNN) interactions. 

The $2N - 3 = 5$ mode stiffnesses $\kappa$ are shown in Fig. \ref{LOfig:fig2}f with increasing stiffness, where we have left out the two translations and one rotation corresponding to  $\kappa=0$ global modes. 
For the experimental data, a $\sim 10^3$ difference in stiffness is visible between the softest mode and the stiffer modes. The modes themselves, obtained from the eigenvectors of $\mathbf{K}$, are plotted in Fig. \ref{LOfig:fig2}g. The softest mode is a shear mode where two opposing particles move inwards, while the other two move outwards, and involves no changes in bond lengths, corresponding thus to the expected floppy mode. The other four normal modes are significantly stiffer, as bonds have to be compressed or extended.

\textbf{Theoretical description.} To better understand the experimental results and to test if all relevant interactions are taken into account, we perform Brownian particle simulations in which all interactions are known, see SI section II\cite{supplementary}. In these simulations, particle neighbors are fixed and connected by springs of stiffness $k_\mathrm{bond} = 50$ $\mathrm{\mu Nm^{-1}}$. 
We solve the overdamped Langevin equation in two dimensions with explicit thermal noise and friction.   
The correct topology is assured by an additional one-sided spring of stiffness \(k_\mathrm{bond}\) that only becomes active when two particles overlap that are not neighbors.
The resulting mode stiffnesses (Fig. \ref{LOfig:fig2}f, labelled `sim') are similar to the experimental results, but for the stiff modes 2-5, the experiments show some variation, which may be caused by variations in bond stiffness caused by differences in DNA concentration \cite{Chakraborty2017}. The overall good match confirms that the system is well approximated as being planar, and as only having normal and no tangential interactions.

\begin{figure*} \centering     \includegraphics[width=\textwidth]{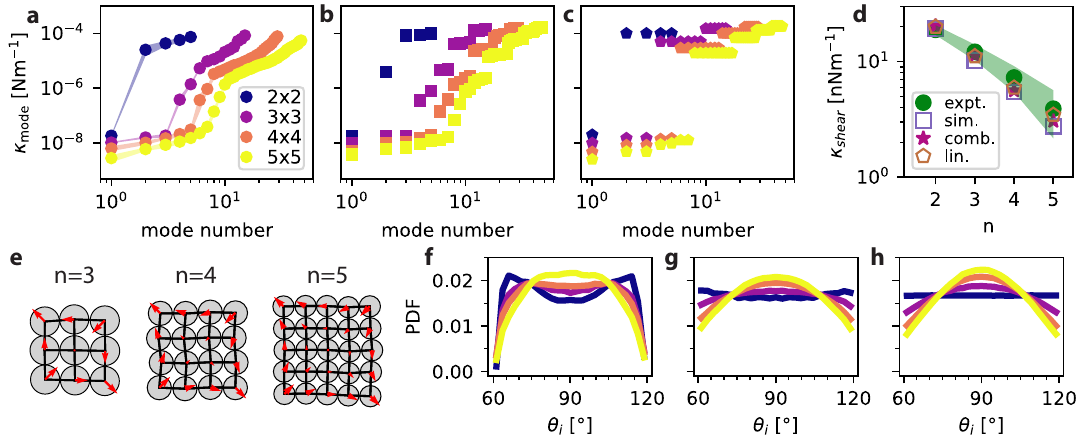}
    \caption{\textbf{Modes for larger systems.} \textbf{a-c)} The mode stiffnesses of larger lattices for \textbf{a)} experiments (expt.), where the line width is equivalent to the standard deviation, \textbf{b)} simulations (sim.), and \textbf{c)} a theoretical map onto linear response (lin.) \textbf{d)} The shear stiffness obtained from the variance of the shear mode projection distribution, where again the experimental standard deviation is indicated. \textbf{e)} Schematics of the softest mode, which is close to a pure shear mode. \textbf{f-h)} The opening angle distributions for \textbf{f)} experiments, \textbf{g)} simulations, and \textbf{h)} the combinatorial approach (comb.). Simulations were done with $k_\mathrm{bond}=$ 50 $\mathrm{\mu Nm^{-1}}$ and the linear response approach (lin.) was done using $k_\mathrm{bond}=$ 50 $\mathrm{\mu Nm^{-1}}$ and the $k_\mathrm{diag.}^n$ derived below. For the linear response data, $\kappa_\mathrm{shear}^\mathrm{lin} = (\sum_\nu \frac{1}{\kappa_\nu}(\vec{m}\cdot\vec{\xi_\nu})^2)^{-1}$ with $\vec{\xi_\nu}$ the eigenvector of $\mathbf{K}$ for mode $\nu$ is plotted in panel e). \label{LOfig:fig3}} \end{figure*}

We now map these results to an effective linear response model. 
Given that we are in the limit of stiff bonds (relative deformations of $<$ 3\%), then for approximately fixed bond length, the conformation of the 2x2 structure can be described by a single opening angle. We assume that this angle has a uniform distribution between the geometrically possible opening angles $\theta = \frac{\pi}{3}$ and $\theta = \frac{2\pi}{3}$, matching reasonably well with experiments \cite{Verweij2023}, see Fig. \ref{LOfig:fig3}f. We can then, to second order, approximate the effective angle stiffness by a normal distribution with the same mean and variance. Finally, the second derivative of the resulting potential with respect to the diagonal distance results in an effective stiffness for a diagonal NNN spring of (see SI section IV\cite{supplementary})
$k_\mathrm{diag.}=\frac{1}{2}\frac{2}{d^2}\frac{k_\mathrm{B}T}{\text{Var}(\theta)} = \frac{108k_\mathrm{B}T}{\pi^2d^2} $, 
where the factor $\frac{1}{2}$ comes from dividing the spring over two diagonals. The resulting spring networks are indicated in Fig. \ref{LOfig:fig1}g.  
For $d=$ 2.12 $\mathrm{\mu m}$ and the above assumptions, $k_\mathrm{diag.}=$ 10 $\mathrm{nNm^{-1}}$. We can then obtain the modes and their stiffnesses by diagonalizing the constructed stiffness matrix (see SI section V \cite{supplementary}) with neigbour stiffness $k_\mathrm{bond}$ and NNN diagonal effective stiffness $k_\mathrm{diag.}$. A similar approach was employed by \cite{mao2013} to derive effective angle springs and low energy modes in a Kagome lattice. We obtain three modes with a stiffness of 2$k_\mathrm{bond}$, one with a stiffness of 2$(k_\mathrm{bond}+k_\mathrm{diag.})$ and one  with stiffness $2k_\mathrm{diag.}$ (`lin.' in Fig. \ref{LOfig:fig2}f). This largely matches experiments and simulations, where one soft mode and four stiff modes are found as well. While the stiff modes acquire a nonlinear component (see below), the stiffness of the soft mode, $20.0$ $\mathrm{nNm^{-1}}$, is very similar to experiments and simulations, both with $19.2$ $\mathrm{nNm^{-1}}$.

\textbf{Larger square lattices.} 
Having obtained a comprehensive understanding of the 2x2 structure, we now turn to larger square lattices with more soft modes. In Fig. \ref{LOfig:fig3}a-c, the mode stiffnesses are shown as a function of mode number for experiments, Brownian simulations, and the linear response approach discussed below, respectively. For our finite systems, Maxwell constraint counting\cite{Maxwell1864} corresponds to $2n^2$ translational degrees of freedom that are countered by $2n^2-2n$ bonds, giving us $2n-3$ floppy modes after taking into account global degrees of freedom. We indeed find 1, 3, 5, and 7 soft modes for, respectively, experimental square lattices with a size of 2x2, 3x3, 4x4, and 5x5.
Furthermore, experiments and simulations show similar mode stiffnesses, although the simulations show groups of near degenerate modes not apparent in experiment. This difference might arise from a variation in DNA concentration and hence bond stiffness in experiments, as mentioned earlier for the 2x2 network.

We find that the softest modes, shown in Fig. \ref{LOfig:fig3}e are nearly pure shear modes. We compute a shear modulus by projecting the experimental trajectories onto a normalised pure shear mode $\vec{m}$, $p(t) = \delta\vec{r}(t)\cdot\vec{m}$.
The distributions $P_n(p)$ for experiments and simulations become more Gaussian with increasing system size as is visible in Fig. S4. We thus use equipartition again to compute $\kappa_{\text{shear}} = k_BT/\text{Var}(p)$, and show the results in Fig. \ref{LOfig:fig3}d. We can derive its value as follows: each angle has a range between $\frac{\pi}{3}$ and $\frac{2\pi}{3}$, but due to entropic effects stemming from the geometric constraints imposed by the square lattice topology, the distribution is not uniform anymore, similar to what we have previously observed for flexible colloidal rings\cite{Verweij2023}. The narrowing of the angle distribution implies that larger lattices assume more square-like configurations due to entropic effects, thereby effectively stabilizing the open structure, in line with predictions \cite{Dennison2013, hu_2018} and experiments on patchy particles.\cite{mao2013} We can obtain $P(\theta_{n\mathrm{x}n})$ using a Monte Carlo approach where we assume fixed bond lengths and we only select sterically allowed angle combinations (see SI section III. \cite{supplementary}) to find angle distributions that closely match experiments and simulations, as shown in Fig. \ref{LOfig:fig3}f-h. Moreover, using this approach we are able to correctly predict the shear distribution $P_n(p)$, see Fig. S4, and the shear modulus, see Fig. \ref{LOfig:fig3}d `comb.' symbols.

To map to linear response, we again introduce NNN springs that derive from the steric constraints, but we need to modify our argument: While there are $(n-1)^2$ floppy squares in the packing, there are only $2n-3$ independent opening angles in the stiff bond limit. To take this effect and the angle distributions into account, we calculate the diagonal spring interaction $k_\mathrm{diag}^n$ for larger systems by scaling $k_\mathrm{diag.}$ as $k_\mathrm{diag.}^n =\frac{2n - 3}{(n-1)^2}\frac{\mathrm{Var}(\theta_{2\mathrm{x}2})}{\mathrm{Var}(\theta_{n\mathrm{x}n})}k_\mathrm{diag.}$.
We show the resulting mode spectrum in Fig. \ref{LOfig:fig3}c.
The soft modes are well-described by the theoretical map onto linear response (`lin.'), but for the slightly stiffer modes, experiments and simulations are influenced by higher order effects due to large lattice deformations (see SI section VI \cite{supplementary}), before reaching the linear response expectation for the stiffest modes. 

In summary, we were able to show that colloidal square lattices made from DNA-functionalized CSLBs can be used as a microscale spring network to study thermally excited normal modes. Brownian particle simulations with a spring stiffness similar to experiments produced similar results. A theoretical map onto linear response with effective NNN springs \cite{Souslov2009} well describes the soft modes and the linear part of the stiff modes. Notably, we find a modified scaling of the shear modulus with system size due to entropic steric effects. Such anomalous scaling of the shear modulus in thermally excited spring networks with excluded volume effects has been observed before \cite{Mao2015,Dennison2013,Zhang2016, hu_2018}, and we can add to the evidence of an entropic origin.
Our experimental model system can be extended to study thermal effects in other more complex ordered and disordered network topologies and investigate the influence of defects. It furthermore provides a starting point to use these effects for designing colloidal structures with global reconfiguration modes with more advanced functionalities, such as colloidal mechanical metamaterials or self-assembled adaptive materials.

DJK gratefully acknowledges funding from the European Research Council (ERC) under the European Union's Horizon 2020 research and innovation program (Grant Agreement No. 758383). SH acknowledges lively discussions with Fred MacKintosh and the hospitality of the Isaac Newton Institute for Mathematical Sciences during the SPL programme, supported by EPSRC grant no EP/R014604/1.

%

\clearpage
\newpage

\begin{widetext}
\begin{center}
\textbf{\huge Soft and stiff normal modes in floppy colloidal square lattices: Supplemental Material}
\end{center}
\end{widetext}

\setcounter{equation}{0}
\setcounter{figure}{0}
\setcounter{table}{0}
\setcounter{section}{0}
\setcounter{page}{1}
\makeatletter
\renewcommand{\theequation}{S\arabic{equation}}
\renewcommand{\thefigure}{S\arabic{figure}}
\renewcommand{\bibnumfmt}[1]{[S#1]}
\renewcommand{\citenumfont}[1]{S#1}


\section{Experimental data acquisition and treatment}
The initial particle positions are obtained from brightfield microscopy videos with a framerate of 20 fps using a least-square fit of a Mie scattering based model implemented in HoloPy \cite{Barkley2020-SI}. This approach has previously been identified to work very well for tracking DNA-functionalized colloid supported lipid bilayers \cite{Verweij2023-SI}. We subtract the two translational and one rotational diffusion mode as follows. The translational diffusion is subtracted by placing the origin of the tracking frame in the center of mass of the networks as
\begin{equation}
    \vec{r}_{i,\mathrm{trans}}(t) = \vec{r}_{i,\mathrm{ini}}(t)
    - \frac{1}{N}\sum_{j=1}^N
    \vec{r}_{j,\mathrm{ini}}(t)
    ,
\end{equation}
\noindent where $\vec{r}_{i,\mathrm{ini}}(t)$ is the initial tracked position of particle $i$ at time $t$ and $N$ is number of particles. The global rotational diffusion mode is subtracted by rotating every particle around the origin with an angle $\theta$, calculated as $\theta (t) = \frac{1}{N}\sum_{i=1}^N\tan^{-1}(\frac{y_{i,\mathrm{trans}}(t)}{x_{i,\mathrm{trans}}(t)})$ where $\big(x_{i,\mathrm{trans}}(t),y_{i,\mathrm{trans}}(t)\big)=\vec{r}_{i,\mathrm{trans}}(t)$. For a particle $i$ in frame $t$, the final translationally and rotationally corrected position is given by 
\begin{equation}
    \vec{r}_{i,\mathrm{rt}}(t) = \vec{r}_{i,\mathrm{trans}}(t)
    \begin{bmatrix}
    \cos{\theta(t)}& \sin{\theta(t)}\\
    -\sin{\theta(t)}&\cos{\theta(t)}
    \end{bmatrix}.
\end{equation}

\begin{table}
\begin{tabular}{ c|c|c|c } 
 Network & \# Expts. & Frames & Time [s] \\
 \hline
 2x2 & 6 & 1.0$\cdot10^5$ & 5.0$\cdot10^3$ \\
 3x3 & 3 & 7.3$\cdot10^4$ & 3.6$\cdot10^3$ \\
 4x4 & 5 & 5.2$\cdot10^4$ & 2.6$\cdot10^3$ \\
 5x5 & 3 & 6.9$\cdot10^4$ & 3.5$\cdot10^3$ \\
\end{tabular}
\caption{Number of experiments, total number of frames and total measurement time per network size $n$.}
\label{table:experiments}
\end{table}

\begin{figure} \centering     \includegraphics[width=5cm]{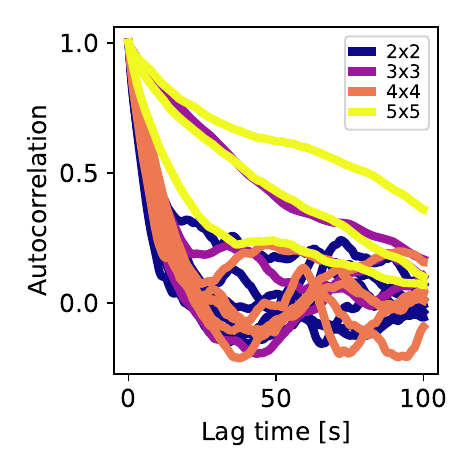}
    \caption{\textbf{Autocorrelation function softest mode.} The autocorrelation as a function of lag time for different network sizes. \label{LOfig:si_fig1}} \end{figure}

An overview of the number of experiments, the imaging time and the number of analyzed frames per experiment for the different network sizes is given in Table \ref{table:experiments}.  For the correlation inversion method to converge to the true dynamical matrix, we need that the ratio $r= T/N_{\text{dof}}\gg 1$, where $T$ is the number of independent samples, and $N_\text{dof} = 2n^2$ is the number of degrees of freedom in the system. However, even at lower $r\gtrsim 1$, the lowest modes and their eigenvalues are well resolved \cite{Henkes12-SI}. As shown in Table \ref{table:experiments}, our number of samples is more than sufficient, with $r$ values in the range from $12500$ for the $2\times2$ to $1380$ for the $5\times5$. However, there are correlations: the longest time scale in our system, that of the autocorrelation of soft modes themselves, is of the order of 10-100 s (see Fig. \ref{LOfig:si_fig1}). Therefore, we expect the effective values of $r$ to be considerably lower. We empirically find reproducible spectra between different experiments of different lengths (Fig. 3a of the main text), and any lack of convergence will be confined to the stiffest end of the spectrum away from the floppy modes \cite{Henkes12-SI,Chen2013-SI}.

\section{Brownian dynamics simulation}
We approximate the system as three dimensional particles moving in two dimensions as the gravitational height $\Delta z$ of one particle is small compared to the diffusion at the time scales we observe. In the simulations, the particle neighbors are fixed and connected by harmonic springs of stiffness $k_\mathrm{bond}$ = 50 $\mathrm{\mu N m^{-1}}$. To assure the correct topology, an additional one-sided harmonic spring of $k_\mathrm{bond}$ is included which becomes active only when particles that are not neighbors are overlapping. 
We solve the overdamped Langevin equation in two dimensions with single particle friction and thermal fluctuations corresponding to the motion of isolated colloids, i.e. Stokes drag sets the single-particle mobility $\mu = 1/(6 \pi \eta R)$ and the translational diffusion $D_T = \mu k_B T$ pairs with it through fluctuation-dissipation. While the true interactions of the colloids with the surface and each other involve hydrodynamics which lowers mobility and hence diffusion our method, unlike microrheology, relies on equilibrium distributions only. Then, any dynamics that follows fluctuation-dissipation and/or equipartition admits the central result $\mathbf{C}_\mathrm{p} = k_\mathrm{B}T \mathbf{K}^{-1}\:$ \cite{Henkes12-SI}. We write the following
equations of motion for particle $i$:
\begin{equation}
\begin{split}
    \dot{x_i} &= \mu \sum_j F_{ij}^x + \eta_x\\
    \dot{y_i} &= \mu \sum_j F_{ij}^y + \eta_y,
\end{split}
\end{equation}
where $\mu = \frac{D_\mathrm{T}}{k_\mathrm{B}T} = 4.0 \cdot 10^7$ $\mathrm{skg^{-1}}$ is the mobility, $F_{ij}^x$ and $F_{ij}^y$ are the spring force components between particles $i$ and $j$ along the $x$- and $y$-axis, respectively, and $\eta_x$ and $\eta_y$ are the thermal noise terms that are drawn from a normal distribution with variance $\sqrt{2D_\mathrm{T}\Delta t} = 1.5 \cdot 10^{-9}$ $\mathrm{m}$ where $\Delta t = 10$ $\mathrm{\mu s}$ is the timestep. A total of $10^9$ steps are performed for each simulation, saving every $10^4$-th step, which corresponds to $10^5$ saved steps that are 0.1 s apart. Simulated autocorrelation times are of the order of 10 times lower than the experimental ones, in line with the estimated change in mobility near the surface \cite{Chakraborty2017-SI}.

\begin{figure} \centering     \includegraphics[width=5cm]{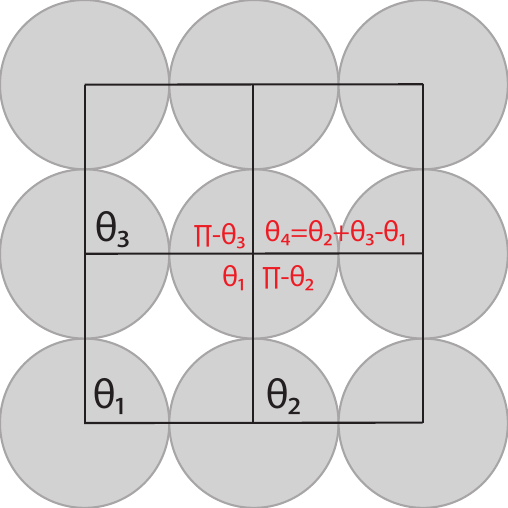}
    \caption{\textbf{Schematic showing an example of a choice of independent angles for a  3x3 network.} In each square, the lower left angle is labeled. For a 3x3 network, only 3 angles can be chosen independently. The fourth angle can be described using the other three, as indicated in red. \label{LOfig:si_fig2}} \end{figure}

\section{Combinatorial approach}
The diagonalization of the theoretical constructed stiffness matrix $\mathbf{K}$ does not give information about the distribution of particle positions. Therefore, we used a combinatorial approach to calculate an additional dataset that provides such distributions. By assuming fixed bond lengths, the square lattices can be fully described by $2n-3$ opening angles $\theta_i$. We then construct a vector $\vec{\theta}_\mathrm{indep.}=\left(\theta_1,\dots,\theta_{2n-3}\right)$ of $2n-3$ independent angles randomly chosen between $\frac{\pi}{3}$ and $\frac{2\pi}{3}$ and repeat this $10^6$ times. Subsequently, only the angle combinations that lead to sterically allowed structures, i.e. structures without overlap, are selected. For example, while a 3x3 network consists of four squares,  there are only three independent angles. If in each square, an angle is defined in the lower left corner as depicted in Fig. \ref{LOfig:si_fig1}, it is straightforward to see that $\theta_1+(\pi-\theta_2)+(\pi-\theta_3)+\theta_4=2\pi$, which leads to $\theta_4 = \theta_2+\theta_3-\theta_1$. The fourth angle can thus be expressed as a combination of the other angles, leaving us with three independent angles. To select the angle combinations that are sterically allowed, we calculated $\theta_4$ and only saved $\vec{\theta}_\mathrm{indep.}$ if $\theta_4>\frac{\pi}{3}$ and $\theta_4<\frac{2\pi}{3}$. 

Since bond lengths are assumed to be fixed, the position of all particles relative to the center of mass can then be given as a function of the independent opening angles, $\vec{r}(\vec{\theta}_\mathrm{indep.})$.
The same mode analysis can be done with this dataset as for experiments and simulations. Instead of time, the position is now a function of the combination of opening angles.

\section{Diagonal interactions}
In this section, a theoretical prediction will be derived for the effective spring constant between next-nearest-neighbors, i.e. particles positioned on the diagonals of the square lattices. First, we consider the 2x2 network. Since we are in the stiff bond limit, we can assume a fixed bond length and therefore can describe the conformation with a single opening angle. This opening angle is assumed to have an uniform angle distribution in line with previous measurements on flat angle distributions for chains of three flexibly linked colloids.\cite{Verweij2021-SI} We then approximate this angle distribution by a normal distribution with the same mean and variance, corresponding to an effective angle spring. Finally, this angle spring is transformed into an effective diagonal spring by taking the second derivative of the potential with respect to the diagonal distance. It should be noted that this method is not linear response, but rather a map onto linear response, for which the stiffness is related to the displacement distributions. \cite{Henkes12-SI}.

To start, we assume a uniform angle probability density function $P(\theta)$ between the geometrically constrained $\frac{\pi}{3}$ and $\frac{2\pi}{3}$ so that $P(\theta)=\frac{3}{\pi}$ for $\frac{\pi}{3}<\theta<\frac{2\pi}{3}$. This distribution has a mean of $\langle\theta\rangle=\int_{\frac{\pi}{3}}^{\frac{2\pi}{3}}\theta P(\theta) d\theta = \frac{\pi}{2}$ and variance of $\left\langle\left(\theta-\langle\theta\rangle\right)^2\right\rangle = \int_{\frac{\pi}{3}}^{\frac{2\pi}{3}}\left(\theta-\langle\theta\rangle\right)^2P(\theta)d\theta=\frac{\pi^2}{108}$. We now approximate this angle distribution with a normal distribution having the same mean and variance and invert the angle variance to obtain an effective angle stiffness $k_{\theta} = \frac{k_\mathrm{B}T}{\left\langle\left(\theta-\langle\theta\rangle\right)^2\right\rangle}=\frac{108}{\pi^2}k_\mathrm{B}T$.

To be able to compare the derived effective stiffness with the mode stiffnesses found in experiments and simulations, we need to get $k_{\theta}$ in the right units by changing the variable. If we have a 2x2 network with particles 1 and 4 on one diagonal and particles 2 and 3 on the other diagonal (as in Fig. \ref{LOfig:fig2}b of the main text) and $\theta$ the opening angle that particles 3, 1, and 2 make, then the length of the diagonals is given by $r_{23} = 2d\sin{\frac{\theta}{2}}$ and $r_{14} = 2d\sin{\frac{\pi-\theta}{2}}$ where the subscripts are the particle indices, $d$ is the particle diameter, $\theta\in\left[\frac{\pi}{3},\frac{2\pi}{3}\right]$, and $\pi-\theta\in\left[\frac{2\pi}{3},\frac{\pi}{3}\right]$. The angle potential is given by $V=\frac{1}{2}k_{\theta}\left(\theta-\frac{\pi}{2}\right)^2$. The effective diagonal stiffness is obtained by taking the second derivative to $r_{23}$ and evaluating around $\frac{\pi}{2}$, so that $k_\mathrm{eff}=\frac{\partial^2 V}{\partial r_{23}^2}\Big|_{\frac{\pi}{2}}$. The first and second derivative with respect to $r_{23}$ are given by
\begin{equation}
\begin{split}
    \frac{\partial V}{\partial r_{23}} =& k_{\theta}\left(\theta-\frac{\pi}{2}\right)\frac{\partial \theta}{\partial r_{23}}\\
    \frac{\partial^2 V}{\partial r_{23}^2} =& k_{\theta}\left(\frac{\partial \theta}{\partial r_{23}}\right)^2 + k_{\theta}\left(\theta - \frac{\pi}{2}\right)\frac{\partial^2 \theta}{\partial r_{23}^2}.
\end{split}
\end{equation}
With $\theta=2\arcsin{\frac{r_{23}}{2d}}$ and evaluating around $\theta=\frac{\pi}{2}$, for which the second term is zero, this leads to
\begin{equation}
    \frac{\partial^2 V}{\partial r_{23}^2}\bigg|_{\theta = \frac{\pi}{2}} = 2\frac{k_{\theta}}{d^2}.
\label{diagonal_stiffness}
\end{equation}

So, the effective stiffness for diagonal interactions is given by $k_\mathrm{diag.,2x2}=\frac{1}{2}\frac{2}{d^2}\frac{k_\mathrm{B}T}{\left\langle\left(\theta-\langle\theta\rangle\right)^2\right\rangle}$, which results in $k_\mathrm{diag.,2x2}=10 \mathrm{nNm^{-1}}$ for $d=2.12$ $\mathrm{\mu m}$, $T=298$ $\mathrm{K}$ and the earlier described angle distribution. The factor $\frac{1}{2}$ comes from the division over two diagonals.

\section{Linear response stiffness matrix}
The theoretical stiffness matrix for a frame with $n\times n$ nodes and $n_b$ bonds (direct bonds and diagonal interactions) can be constructed following ref.\cite{Lubensky2015-SI}. First, the $n_b$x$2n$ compatibility matrix $\mathbf{C}$ is constructed that maps the bond elongation vector $\vec{e}$ to the particle displacement vector $\vec{u}$ as $\vec{e}=\mathbf{C}\vec{u}$. From this, the $2n$x$2n$ stiffness matrix $\mathbf{K}$ is obtained as $\mathbf{K}=\mathbf{C}\vec{k}\mathbf{C^T}$, where $\vec{k}$ is the vector containing the stiffnesses for the bonds and the diagonal interactions ($k_\mathrm{bond}$ and $k_\mathrm{diag.}$, respectively).

For a 2x2 structure with bonds 1-2 and 3-4 in the $x$-direction and bonds 1-3 and 2-4 in the $y$-direction (same as in Fig. \ref{LOfig:fig2}b of the main text), the theoretical linear response stiffness matrix $\mathbf{K}$ is given by

\begin{equation*}
\mathbf{K} = 
\begin{bmatrix}
c & b & -a & 0 & 0 & 0 & -b & -b \\
b & c & 0 & 0 & 0 & -a & -b & -b\\
-a & 0 & c & -b & -b & b & 0 & 0\\
0 & 0 & -b & c & b & -b & 0 & -a\\
0 & 0 & -b & b & c & -b & -a & 0\\
0 & -a & b & -b & -b & c & 0 & 0\\
-b & -b & 0 & 0 & -a & 0 & c & b\\
-b & -b & 0 & -a & 0 & 0 & b & c
\end{bmatrix},
\end{equation*}
with $a = k_\mathrm{bond}$, $b = \frac{k_\mathrm{diag.}}{2}$, and $c = k_\mathrm{bond} + \frac{k_\mathrm{diag.}}{2}$.

\begin{figure}[H] \centering     \includegraphics[width=5cm]{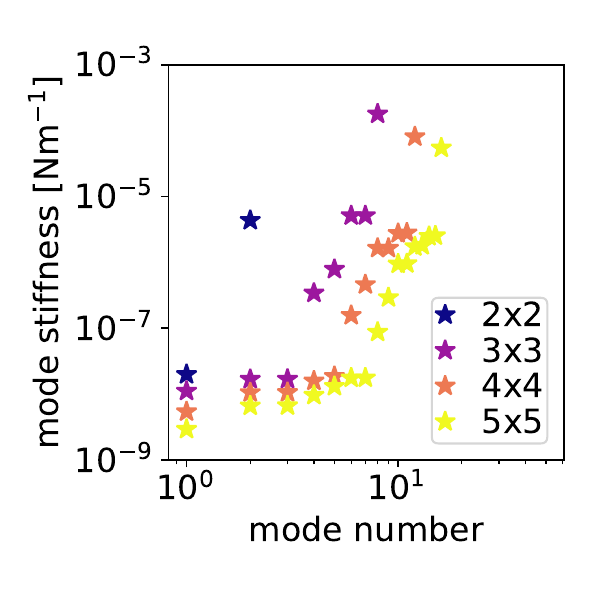}
    \caption{\textbf{Nonlinear effects from soft modes.} Mode stiffnesses as a function of mode number for the combinatorial approach showing more modes with a finite stiffness than just the soft modes. \label{LOfig:si_fig3}} \end{figure}

\begin{figure}[H] \centering     \includegraphics[width=9cm]{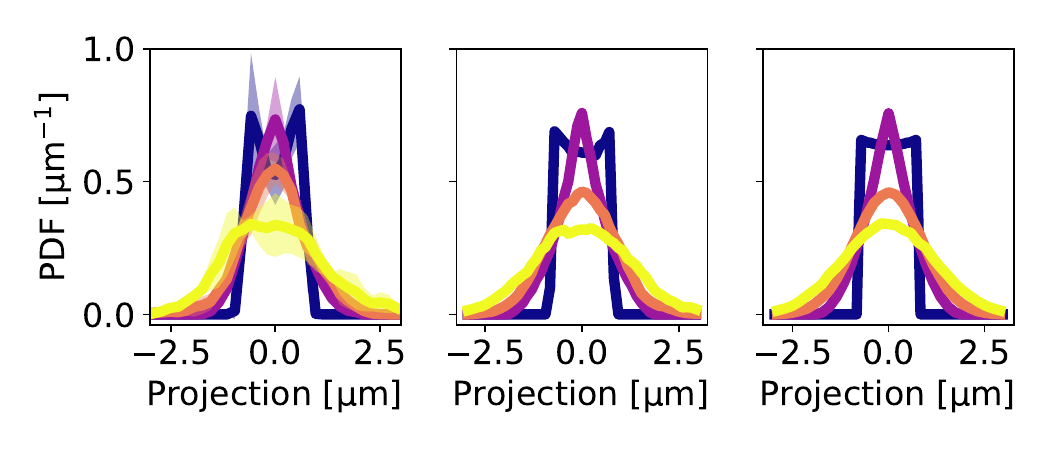}
    \caption{\textbf{Shear mode projection distribution} The distributions of the uniform shear modes of the different lattice sizes for (from left to right) experiments, simulations, and the combinatorial approach, respectively. \label{LOfig:si_fig4}} \end{figure}

\section{Nonlinear effects}
In Fig. \ref{LOfig:fig3}a-b of the main text, the stiffer modes appear softer than expected based on the theoretical linear response. This is a nonlinear effect stemming from the large displacements of the soft modes. A similar plot is shown in Fig. \ref{LOfig:si_fig3} where the mode stiffnesses for the combinatorial approach are shown. In linear response, these should just contain the soft modes and not the modes where bond lengths are changed. However, the large displacements along the soft modes result in some effective projections onto stiffer modes, resulting in more modes than just the soft modes. This is likely the reason why some stiff mode stiffnesses for experiments and simulations are shifted to lower values and some smaller values show up in the 2x2 stiffness matrix shown in Fig. \ref{LOfig:fig2}e of the main text.

\section{Shear mode projection distribution}
In Fig. \ref{LOfig:si_fig4} the projection distributions of respectively experiments, simulations and the combinatorial approach are shown which were used to obtain the shear moduli reported in the main text. The 2x2 network projections have a nearly uniform distribution, while the larger network projections resemble a normal distribution more closely.

\end{document}